\documentclass{revtex4-2}
\usepackage[utf8]{inputenc}
\usepackage{amsbsy}
\usepackage{amsopn}
\usepackage{amstext}
\usepackage{amsmath,amsthm,amsfonts,amssymb}
\usepackage[mathcal]{eucal}
\usepackage{mathrsfs}
\usepackage{braket}
\usepackage[english]{babel}
\usepackage{color}
\usepackage{esint}
\usepackage{graphicx}
\usepackage{float}
\usepackage{units}
\usepackage{blindtext}
\usepackage{hyperref}
\usepackage{textcomp}
\usepackage{orcidlink}
\usepackage{lineno}

\DeclareGraphicsExtensions{.png,PNG,.pdf,.PDF}
\usepackage{hyperref}
\usepackage{slashed}
\newcommand{\ie}{\begin{equation}}
\newcommand{\fe}{\end{equation}}
\newcommand{\se}{\begin{eqnarray}}
\newcommand{\ff}{\end{eqnarray}}
\begin{document}

\title{Comment on ``Neutrino interaction with matter in a noninertial frame''}
\author{R. R. S. Oliveira\,\orcidlink{0000-0002-6346-0720}}
\email{rubensrso@fisica.ufc.br}
\affiliation{Departamento de F\'isica, Universidade Federal da Para\'iba, Caixa Postal 5008, 58051-900, Jo\~ao Pessoa, PB, Brazil}

%%%%%%%%%%%%%%%%%%%%%%%%%%%%%%%%%%%%%%%%%%%%%%%%%%%%%%%%%%%%%%%%%%%%%%

\date{\today}

\begin{abstract}

In this comment, we obtain the complete energy levels for Dvornikov's paper \cite{Dvornikov}, that is, the energy levels dependent on two quantum numbers, namely, the radial quantum number (given by $N$) and the angular quantum number (given by $J_z$). In particular, what motivated us to do this was the fact that the quantized energy levels for particles (fermions or bosons) in polar, cylindrical, or spherical coordinates depend on two quantum numbers: a radial quantum number and an angular quantum number. From this, the following question/doubt arose: Why do the energy levels in Dvornikov's paper only depend on one quantum number? That is, Where did the angular quantum number given by $J_z$ go? So, using Studenikin's paper \cite{Studenikin} as a starting point (as well as others in the literature), we write one of the equations from Dvornikov's paper \cite{Dvornikov} in matrix form. Next, we use the four-component Dirac spinor and obtain a set/system of four coupled first-order differential equations. From the first two equations with $m\to 0$ (massless neutrino or ultrarelativistic regime), we obtain a (compact) second-order differential equation for the last two spinor components. So, solving this equation, we obtain the neutrino energy levels, which explicitly depend on both $N$ and $J_z$. Finally, we note that for $J_z>0$ (positive angular momentum) with $u=+1$ (component $\psi_3$), we obtain exactly the particular energy levels of Dvornikov's paper \cite{Dvornikov}.

\end{abstract}

\maketitle

%%%%%%%%%%%%%%%%%%%%%%%%%%%%%%%%%%%%%%%%%%%%%%%%%%%%%%%%%%%%%%%%%%%%%%%%%%%%%%%%%%
\section{Introduction}

In a paper published in the Journal of High Energy Physics (JHEP), entitled ``Neutrino interaction with matter in a noninertial frame'', Dvornikov \cite{Dvornikov} studied the system of massive and mixed neutrinos interacting with background matter moving with an acceleration. To do this study, he worked with the Dirac equation for a single neutrino in the noninertial frame with constant angular velocity in cylindrical coordinates, where he used the vierbein vectors formalism to write the Dirac equation in a curved space-time. After solving the Dirac equation (transformed into an effective Schrödinger equation) for $m\to 0$ (massless neutrino or ultrarelativistic regime), he found the neutrino energy levels for ultrarelativistic particles propagating in the rotating matter, where such levels are quantized in terms of a quantum number $N$, whose values are $N=0,1,2,\ldots$ (and arises in the solution of a second-order radial differential equation, i.e. such quantum number also can be called of radial quantum number). Besides, he obtained the resonance condition for neutrino oscillations and examined how they can be affected by matter rotation. Finally, he also compared the results with the findings of other authors who have studied analogous problems previously. In particular, this paper is well-written, very interesting, and covers a very important subject involving Dirac neutrinos in rotating frames. Furthermore, the formalism used by him is also of great relevance in the literature when working with Dirac fermions in curved space-times or rotating frames \cite{Bakke,Bakke2,Bakke3,Bakke4,Cuzinatto,O1,O2,O3,O4,O5,O6,Andrade,Cunha,Bragança}.

However, starting from the fact that the quantized energy levels for particles (fermions or bosons) in polar, cylindrical, or spherical coordinates (i.e. 2D and 3D systems, or (2+1)D and (3+1)D space-times) depend on two quantum numbers (a radial quantum number and an angular quantum number) \cite{Bakke,Bakke2,Bakke3,Bakke4,Cuzinatto,O2,O3,O4,O5,O6,Andrade,Cunha,Bragança,Carvalho,Cuzinatto2,Castro}, the following question/doubt arose: Why do the energy levels in Dvornikov's paper only depend on one quantum number? In other words, Where did the angular quantum number given by $J_z$ go? So, as this was not explicit in Dvornikov's paper (since $J_z\neq 0$) \cite{Dvornikov}, this motivated us to look for such a quantum number, or better, to obtain the energy levels in their general/complete form (such as happens for fermions or bosons in \cite{Bakke,Bakke2,Bakke3,Bakke4,Cuzinatto,O2,O3,O4,O5,O6,Andrade,Cunha,Bragança,Carvalho,Cuzinatto2,Castro}). Therefore, the goal of the present comment is to obtain in detail the complete energy levels for Dvornikov's paper \cite{Dvornikov}, that is, the energy levels dependent on two quantum numbers (radial and angular quantum numbers). To achieve this, we will use Studenikin's paper \cite{Studenikin} as a starting point (as well as Refs. \cite{Bakke,Bakke2,Bakke3,Bakke4,Cuzinatto,O1,O2,O3,O4,O5,O6,Cunha,Bragança}). In this paper, Studinenk \cite{Studenikin} (section 2.4: Neutrino quantum states in rotating medium) obtained from the (3+1)-dimensional Dirac equation in Cartesian coordinates a set/system of four coupled first-order differential equations (i.e. each equation contained three of the four spinor components). Considering $m \simeq 0$ (massless neutrinos or with a very small mass, or kinetic term + interaction term $\gg$ mass term), he reduced the system into two independent sets of two equations, that couple the components in pairs: ($\Psi_1,\Psi_2$) and ($\Psi_3,\Psi_4$). So, from the pair ($\Psi_1,\Psi_2$), which contains a matter term (or interaction term), he obtained the quantized energy levels of the neutrino. Last but not least, it is interesting to mention that this author also has some papers published together with Dvornikov \cite{Dvornikov2,Dvornikov3,Dvornikov4,Dvornikov5}, that is, both are experts in the study of Dirac neutrinos (or Dirac equation for neutrinos interacting with background matter).

%-------------------------------------------------------------------

\section{Quick review of the main steps that Dvornikov took to obtain the energy levels}

According to Dvornikov \cite{Dvornikov}, it is known that the motion of a test particle in a noninertial frame is equivalent to the interaction of this particle with a gravitational field (in fact, this is a consequence of Einstein's equivalence principle). In this way, the Dirac equation for a massive neutrino moving in a (3+1)D curved space-time and interacting with background matter is given by the following equation \cite{Dvornikov}:
\begin{equation}\label{1}
[i\gamma^\mu (x)\nabla_\mu-m]\psi=\frac{1}{2}\gamma_\mu (x)g^\mu [1-\gamma^5 (x)]\psi,
\end{equation}
where $\gamma^\mu (x)=e^{\ \mu}_a \gamma^a$ and $\gamma_\mu (x)=e^a_{\ \mu}\gamma_a$ are the curved gamma matrices, $\gamma^a$ and $\gamma_a$ are the flat gamma matrices, $e^{\ \mu}_a$ is the vierbein and and $e^a_{\ \mu}$ is its inverse (and must satisfy the orthogonality condition given by $e^a_{\ \mu} e^{\ \mu}_b =\delta^a_b$), $\nabla_\mu=\partial_\mu+\Gamma_\mu$ is the covariant derivative, $\Gamma_\mu=-\frac{i}{4}\sigma^{ab}\omega_{ab\mu}=-\frac{i}{4}\sigma^{ab}e^{\ \nu}_a e_{b\nu;\mu}$ is the spin connection, $\sigma^{ab}=\frac{i}{2}[\gamma^a,\gamma^b]$ are the generators of the Lorentz transformations in a locally Minkowskian frame, $\gamma^5 (x)=-\frac{i}{4!}E^{\mu\nu\alpha\beta}\gamma_\mu (x)\gamma_\nu (x)\gamma_\alpha (x) \gamma_\beta (x)$ is the curved fifth gamma matrix, $E^{\mu\nu\alpha\beta}=\frac{1}{\sqrt{-g}}\varepsilon^{\mu\nu\alpha\beta}$ is the covariant antisymmetric tensor in curved space-time, $g$ is the determinant of the metric ($g=$ det$(g_{\mu\nu})$), $g^\mu$ is the diagonal effective potential in the mass eigenstates basis (i.e. the last term of Eq. \eqref{1} is the background matter or simply the matter term), and $\psi$ is the four-component Dirac spinor, respectively. 

According to Dvornikov \cite{Dvornikov}, the interval (line element) in the rotating frame is given by:
\begin{equation}\label{interval1}
ds^2=g_{\mu\nu}dx^\mu dx^\nu=(1-\omega^2 r^2)dt^2-dr^2-2\omega r^2 dt d\phi-r^2 d\phi^2-dz^2,
\end{equation}
where $x^\mu=(t,r,\phi,z)$ is the position four-vector in cylindrical coordinates, $\omega$ is the constant angular velocity, and $g_{\mu\nu}$ is the metric tensor, respectively. Besides, one can check that the metric tensor in Eq. \eqref{interval1} can be diagonalized (with $\eta_{ab}=e_a^{\ \mu}e_b^{\ \nu}g_{\mu\nu}$) if we use the following vierbein vectors \cite{Dvornikov}:
\begin{eqnarray}\label{vierbein1}
&& e_0^{\ \mu}=\left(\frac{1}{\sqrt{1-\omega^2 r^2}},0,0,0\right),
\nonumber\\
&& e_1^{\ \mu}=\left(0,1,0,0\right),
\nonumber\\
&& e_2^{\ \mu}=\left(\frac{\omega r}{\sqrt{1-\omega^2 r^2}},0,\frac{\sqrt{1-\omega^2 r^2}}{r},0\right),
\nonumber\\
&& e_3^{\ \mu}=(0,0,0,1),
\end{eqnarray}
where $\eta_{ab}$=diag$(+1,-1,-1,-1)$ is the metric in a locally Minkowskian frame. Besides, the (non-zero) components of the connection one-form $\omega_{ab}=\omega_{ab\mu}dx^\mu$ are given as follows \cite{Dvornikov}:
\begin{eqnarray}\label{connection}
&& \omega_{01\mu}=-\omega_{10\mu}=\left(-\frac{\omega^2 r}{\sqrt{1-\omega^2 r^2}},0,-\frac{\omega r}{\sqrt{1-\omega^2 r^2}},0\right),
\nonumber\\
&& \omega_{02\mu}=-\omega_{20\mu}=\left(0,\frac{\omega}{1-\omega^2 r^2},0,0\right),
\nonumber\\
&& \omega_{12\mu}=-\omega_{21\mu}=\left(\frac{\omega}{\sqrt{1-\omega^2 r^2}},0,\frac{1}{\sqrt{1-\omega^2 r^2}},0\right),
\end{eqnarray}
where implies in the following (non-zero) components of the spin connection $\Gamma_\mu=\frac{1}{8}[\gamma^a,\gamma^b]\omega_{ab\mu}=\frac{1}{4}\gamma^a\gamma^b\omega_{ab\mu}$ (with $\gamma^a\gamma^b=-\gamma^b\gamma^a$ and $a\neq b$):
\begin{align}
\Gamma_t
&= \frac{1}{4}\gamma^a\gamma^b\omega_{ab t},  \nonumber\\
&= \frac{1}{4}\gamma^0\gamma^1\omega_{01t}+\frac{1}{4}\gamma^1\gamma^0\omega_{10t}+\frac{1}{4}\gamma^1\gamma^2\omega_{12t}+\frac{1}{4}\gamma^2\gamma^1\omega_{21t}, 
\nonumber\\
&= \frac{1}{2}\gamma^0\gamma^1\omega_{01t}+\frac{1}{2}\gamma^1\gamma^2\omega_{12t},
\nonumber\\
&= -\frac{\omega^2 r}{2\sqrt{1-\omega^2 r^2}}\gamma^0\gamma^1+\frac{\omega}{2\sqrt{1-\omega^2 r^2}}\gamma^1\gamma^2,
\nonumber\\
\Gamma_r
&= \frac{1}{4}\gamma^a\gamma^b\omega_{ab r},  \nonumber\\
&= \frac{1}{4}\gamma^0\gamma^2\omega_{02r}+\frac{1}{4}\gamma^2\gamma^0\omega_{20r}, 
\nonumber\\
&= \frac{1}{2}\gamma^0\gamma^2\omega_{02r},
\nonumber\\
&= \frac{\omega}{2(1-\omega^2 r^2)}\gamma^0\gamma^2,
\nonumber\\
\Gamma_\phi
&= \frac{1}{4}\gamma^a\gamma^b\omega_{ab \phi},  \nonumber\\
&= \frac{1}{4}\gamma^0\gamma^1\omega_{01\phi}+\frac{1}{4}\gamma^1\gamma^0\omega_{10\phi}+\frac{1}{4}\gamma^1\gamma^2\omega_{12\phi}+\frac{1}{4}\gamma^2\gamma^1\omega_{21\phi}, 
\nonumber\\
&= \frac{1}{2}\gamma^0\gamma^1\omega_{01\phi}+\frac{1}{2}\gamma^1\gamma^2\omega_{12\phi},
\nonumber\\
&= -\frac{\omega r}{2\sqrt{1-\omega^2 r^2}}\gamma^0\gamma^1+\frac{1}{2\sqrt{1-\omega^2 r^2}}\gamma^1\gamma^2=\frac{\Gamma_t}{\omega}.
\end{align}

In particular, the components above are in perfect agreement with \cite{Bakke} (cited by Dvornikov \cite{Dvornikov}) as well as with \cite{Bakke2,Bakke3,Cuzinatto,O1,O2,O3,O4,O5}. With this, Dvornikov \cite{Dvornikov} rewrote Eq. \eqref{1} in the following equation (with $g^\mu=(g^t,0,0,0)=(g^0,0,0,0)$ and $\gamma^5(x)=\gamma^5=\gamma_5=i\gamma^0\gamma^1\gamma^2\gamma^3$):
\begin{equation}\label{2}
[\mathcal{D}-m]\psi=\frac{1}{2}\sqrt{1-\omega^2 r^2}\gamma^0 g^0(1-\gamma^5)\psi,
\end{equation}
where
\begin{equation}\label{3}
\mathcal{D}=i\frac{\gamma^0+\omega r\gamma^2}{\sqrt{1-\omega^2 r^2}}\partial_0+i\gamma^1 \left(\partial_r+\frac{1}{2r}\right)+i\gamma^2\frac{\sqrt{1-\omega^2 r^2}}{r}\partial_\phi+i\gamma^3 \partial_z-\frac{\omega}{2(1-\omega^2 r^2)}\gamma^3\gamma^5,
\end{equation}
being the matrices $\gamma^0$, $\gamma^k$ $(k=1,2,3)$, and $\gamma^5$ written as (chiral representation):
\begin{equation}\label{matrices}
\gamma^0=\left(
    \begin{array}{cc}
      0\ &  -1 \\
      -1\ & 0 \\
    \end{array}
  \right), \ \  \gamma^k=\left(
    \begin{array}{cc}
      0 & \sigma^k \\
      -\sigma^k & \ 0 \\
    \end{array}
  \right), \ \  \gamma^5=\left(
    \begin{array}{cc}
      1 & \ 0 \\
      0 & -1 \\
    \end{array}
  \right).
\end{equation}

So, considering the following spinor:
\begin{equation}\label{spinor}
\psi=\text{exp}(-iEt+iJ_z\phi+ip_z z)\psi_r, 
\end{equation}
as well as a slow rotation regime ($\omega\ll 1$ or $\omega r\ll 1$), Dvornikov \cite{Dvornikov} obtained:
\begin{equation}\label{4}
\left[i\gamma^1\left(\partial_r+\frac{1}{2r}\right)-\gamma^2\left(\frac{J_z}{r}-\omega r E\right)+\gamma^0\left(E-\frac{g^0}{2}\right)-\gamma^3 p_z+\frac{g^0}{2}\gamma^0\gamma^5-\frac{\omega}{2}\gamma^3\gamma^5-m\right]\psi_r=0,
\end{equation}
where $\psi_r$ the spinor depending on the radial coordinate, and $J_z=\frac{1}{2}-l=\pm\frac{1}{2},\pm\frac{3}{2},\pm\frac{5}{2},\ldots$ (with $l=0,\pm 1,\pm 2,\ldots$) is the angular quantum number (i.e. a quantum number associated with the angular part of the spinor). Besides, Dvornikov \cite{Dvornikov} considered a transformation (equivalent to a boost) in the wave function $\psi_r$ given by:
\begin{equation}
\psi=U_3\Tilde{\psi}_r, \ \ U_3=\text{exp}\left(\frac{\upsilon}{2}\gamma^0\gamma^3\right)=\cosh{\frac{\upsilon}{2}}+\gamma^0\gamma^3\sinh{\frac{\upsilon}{2}}, \ \ \upsilon=\frac{\omega}{g^0},
\end{equation}
where the spinor $\tilde{\psi}_r$ obeys the following equation \cite{Dvornikov}:
\begin{equation}\label{5}
\left[\gamma^a Q_a+\frac{\tilde{g}^0}{2}\gamma^0\gamma^5-m\right]\tilde{\psi}_r=0,
\end{equation}
being $Q_a=\left(\tilde{E},-i(\partial_r+\frac{1}{2r}), (\frac{J_z}{r}-\omega r E),\tilde{p}_z\right)$, $\tilde{E}=E'\cosh{\upsilon}-p_z\sinh{\upsilon}$ ($E'=E-\frac{g^0}{2}$), $\tilde{p}_z=p_z \cosh{\upsilon}-E'\sinh{\upsilon}$, and $\tilde{g}^0=g^0\cosh{\upsilon}-\omega\sinh{\upsilon}$, respectively. Besides, Dvornikov \cite{Dvornikov} also considered the following relation for $\Tilde{\psi}_r$:
\begin{equation}\label{relation}
\Tilde{\psi}_r=\Pi\Phi, \ \ \Pi=\gamma^a Q_a-\frac{\tilde{g}^0}{2}\gamma^0\gamma^5+m,
\end{equation}
where $\Phi=\Phi(r)$ is a new spinor. With respect to this relation, it aims to arrive at a ``quadratic Dirac equation'', that is, arriving at a second-order differential equation without passing (or using) directly through the spinor components, or better, without starting from the first-order differential equations coupled with the spinor components, such as is done in Refs. \cite{Bakke,Bakke2,Bakke3,Bakke4,Cuzinatto,O1,O2,O3,O4,O5,O6,Cunha,Bragança}. In particular, this method for obtaining a ``quadratic Dirac equation'' also has a lot of relevance in the literature, where several papers have already been made based on it \cite{Andrade,Gavrilov1,Gavrilov2,Vakarchuk,Peres,O7,O8,O9,O10}.

Therefore, using \eqref{relation}, Dvornikov \cite{Dvornikov} obtained the following second-order differential equation (or ``quadratic Dirac equation''):
\begin{equation}\label{6}
\left[\partial_r^2+\frac{1}{r}\partial_r-\frac{1}{4r^2}+\tilde{E}^2-\tilde{p}^2_z-\left(\frac{J_z}{r}-\omega r E\right)^2-m^2+\frac{(\tilde{g}^0)^2}{4}+\left(E\omega+\frac{J_z}{r^2}\right)\Sigma_3-\tilde{g}^0\tilde{E}\gamma^5+m\tilde{g}^0\gamma^0\gamma^5\right]\Phi=0.
\end{equation}

So, according to Dvornikov \cite{Dvornikov}, Eq. \eqref{6} can be solved for ultrarelativistic neutrinos when the particle mass is neglected ($m\to 0$). In this case, we represent $\Phi=\upsilon\varphi$, where $\upsilon$ is a constant spinor (and satisfies $\Sigma_3\upsilon=\sigma\upsilon$ and $\gamma^5\upsilon=\chi\upsilon$, where $\sigma=\chi=\pm 1$) and $\varphi=\varphi (r)$ is a scalar function \cite{Dvornikov}. In addition, introducing a new variable given by $\rho=E\omega r^2$, Dvornikov \cite{Dvornikov} obtained the following equation:
\begin{equation}\label{7}
\left[\rho\frac{d^2}{d\rho}+\frac{d}{d\rho}-\frac{1}{4\rho}\left(l+\frac{\sigma-1}{2}\right)^2-\frac{\rho}{4}-\frac{1}{2}\left(l-\frac{\sigma+1}{2}\right)+\kappa\right]\varphi_\sigma=0,
\end{equation}
where 
\begin{equation}\label{k}
\kappa=\frac{1}{4E\omega}\left(\tilde{E}^2-\tilde{p}^2_z+\frac{(\tilde{g}^0)^2}{4}-\tilde{g}^0\tilde{E}\chi\right).
\end{equation}

According to Dvornikov \cite{Dvornikov}, the solution of the equation above (which vanishes at infinity) depends on the associated Laguerre polynomials; consequently, the parameter $\kappa$ must be equal to a positive integer, that is, $\kappa=N$, where $N=0,1,2,\ldots$ is the radial quantum number (i.e. a quantum number associated with the radial/spatial part of the spinor). Therefore, using the condition $\kappa=N$, Dvornikov \cite{Dvornikov} obtained the following relativistic energy levels (or high energy spectrum) for the Dirac neutrino:
\begin{align}\label{spectrum1}
& [E_A-2N\omega-g^0]^2=(2N\omega)^2+4N\omega g^0+\left(p_z-\frac{\omega}{2}\right)^2,  
\nonumber\\
& [E_S-2N\omega]^2=(2N\omega)^2+\left(p_z+\frac{\omega}{2}\right)^2,
\end{align}
where $E_A=E(\chi=+1)$ and $E_S=E(\chi=-1)$ are the energies of active and sterile neutrinos, respectively. From the above, we clearly see that the Dvornikov energy levels do not depend (explicitly or implicitly) on the angular quantum number $J_z$ (or $l$), which is quite strange since it should depend on him too (such as happens in \cite{Bakke,Bakke2,Bakke3,Bakke4,Cuzinatto,O2,O3,O4,O5,O6,Andrade,Cunha,Bragança,Carvalho,Cuzinatto2,Castro}).

%-------------------------------------------------------------------

\section{The complete energy levels for Dvornikov’s paper}

Here, we will obtain the complete energy levels for Dvornikov's paper, that is, the energy levels dependent on two quantum numbers: the radial quantum number $N$ and the angular quantum number $J_z$. For this, we will follow the same procedure done by \cite{Studenikin} (section 2.4: Neutrino quantum states in rotating medium), and in a way, such as was done in Refs. \cite{Bakke,Bakke2,Bakke3,Bakke4,Cuzinatto,O1,O2,O3,O4,O5,O6,Andrade,Cunha,Bragança} (that is, we will follow a simpler path than Dvornikov \cite{Dvornikov} did). So, first of all, we need to find the matrix form of Eq. \eqref{4}, which is easily found using the matrices \eqref{matrices} and the form of the Pauli matrices. However, knowing that the space-time (line element or metric) worked by Dvornikov \cite{Dvornikov} has a signature $(+,-,-,-)$, or a flat metric $\eta_{ab}=\eta^{ab}$=diag$(+1,-1,-1,-1)$, implies that: $\gamma^a=\eta^{ab}\gamma_{b}=(\gamma^0,\gamma^1,\gamma^2,\gamma^3)=(\gamma_0,-\gamma_1,-\gamma_2,-\gamma_3)$, that is, go up or down the spatial indexes get a minus sign ($\gamma^i=-\gamma_i$ or $\gamma_i=-\gamma^i$, with $i=1,2,3$) \cite{Greiner,Strange}. Therefore, using this information and the matrices \eqref{matrices} together with the following Pauli matrices:
\begin{equation}\label{Paulimatrices}
 \sigma_1=\sigma_x=\left(
    \begin{array}{cc}
      0\ &  1 \\
      1\ & 0 \\
    \end{array}
  \right), \ \  \sigma_2=\sigma_y=\left(
    \begin{array}{cc}
      0 & -i \\
      i & \ 0 \\
    \end{array}
  \right), \ \
  \sigma_3=\sigma_z=\left(
    \begin{array}{cc}
      1 & \ 0 \\
      0 & -1 \\
    \end{array}
  \right),
\end{equation}
Eq. \eqref{4} will be written in the following matrix form:
\begin{equation}\label{8}
\left(\begin{array}{cccc}
-m & 0 & g^0-E+p_z-\frac{\omega}{2} & -i\partial_r-\frac{i}{2r}-i\frac{J_z}{r}+i\omega r E \\
0 & -m & -i\partial_r-\frac{i}{2r}+i\frac{J_z}{r}-i\omega r E & g^0-E-p_z+\frac{\omega}{2} \\
-E-p_z-\frac{\omega}{2} & i\partial_r+\frac{i}{2r}+i\frac{J_z}{r}-i\omega E r & -m & 0	\\
i\partial_r+\frac{i}{2r}-i\frac{J_z}{r}+i\omega r E & -E+p_z+\frac{\omega}{2} & 0 & -m
\end{array}\right)\psi_r=0.
\end{equation}

Now, writing the radial spinor $\psi_r$ such as:
\begin{equation}\label{spinor2}
\psi_r(r)=\left(\begin{array}{c}
\psi_1(r) \\
\psi_2(r) \\
\psi_3(r) \\
\psi_4(r)
\end{array}\right), 
\end{equation}
we can obtain from \eqref{8} a set/system of four coupled first-order differential equations, given by:
\begin{align}
& \left(g^0-E+p_z-\frac{\omega}{2}\right)\psi_3(r)-i\left(\partial_r+\frac{1}{2r}+\frac{J_z}{r}-\omega E r\right)\psi_4(r)=m\psi_1(r),
\nonumber\\
& \left(g^0-E-p_z+\frac{\omega}{2}\right)\psi_4(r)-i\left(\partial_r+\frac{1}{2r}-\frac{J_z}{r}+\omega E r\right)\psi_3(r)=m\psi_2(r),
\nonumber\\
& \left(-E-p_z-\frac{\omega}{2}\right)\psi_1(r)+i\left(\partial_r+\frac{1}{2r}+\frac{J_z}{r}-\omega E r\right)\psi_2(r)=m\psi_3(r),
\nonumber\\
& \left(-E+p_z+\frac{\omega}{2}\right)\psi_2(r)+i\left(\partial_r+\frac{1}{2r}-\frac{J_z}{r}+\omega E r\right)\psi_1(r)=m\psi_4(r).
\end{align}

So, for $m\to 0$ (or $m\simeq 0$) \cite{Studenikin,Dvornikov}, we have:
\begin{align}
& \left(g^0-E+p_z-\frac{\omega}{2}\right)\psi_3(r)-i\left(\partial_r+\frac{1}{2r}+\frac{J_z}{r}-\omega E r\right)\psi_4(r)=0,
\nonumber\\
& \left(g^0-E-p_z+\frac{\omega}{2}\right)\psi_4(r)-i\left(\partial_r+\frac{1}{2r}-\frac{J_z}{r}+\omega E r\right)\psi_3(r)=0,
\nonumber\\
& \left(-E-p_z-\frac{\omega}{2}\right)\psi_1(r)+i\left(\partial_r+\frac{1}{2r}+\frac{J_z}{r}-\omega E r\right)\psi_2(r)=0,
\nonumber\\
& \left(-E+p_z+\frac{\omega}{2}\right)\psi_2(r)+i\left(\partial_r+\frac{1}{2r}-\frac{J_z}{r}+\omega E r\right)\psi_1(r)=0.
\end{align}

In particular, we see that both/all equations above have the contribution of $\omega$ (noninertial/rotation term); however, only the first two have the contribution of $g^0$ (matter term). So, as our goal is to get the energy levels written as a function of both $\omega$ and $g^0$, it implies that the first two equations are the ones that we will work on. In this way, isolating $\psi_4$ in the second equation and substituting it into the first (i.e. decoupling the third component), as well as isolating $\psi_3$ in the first equation and substituting it into the second (i.e. decoupling the fourth component), we obtain the following second-order differential equation (in a compact form) for $\psi_3$ and $\psi_4$ (such as is also done in Refs. \cite{Bakke,Bakke2,Bakke3,Bakke4,Cuzinatto,O1,O2,O3,O4,O5,O6,Cunha,Bragança}):
\begin{equation}\label{9}
\left[\partial_r^2+\frac{1}{r}\partial_r-\frac{(J_z-\frac{u}{2})^2}{r^2}-(\omega r E)^2+(E-g^0)^2-\left(p_z-\frac{\omega}{2}\right)^2+2\omega E\left(J_z+\frac{u}{2}\right)\right]\psi_u(r)=0, \ \ (u=\pm 1),
\end{equation}
where $\psi_+(r)=\psi_3(r)$ and $\psi_-(r)=\psi_4(r)$.

In addition, introducing a new variable given by $\rho=E\omega r^2$ \cite{Dvornikov} ($E=E_{particle}>0$ or $E=E_{antiparticle}>0$ \cite{Greiner}), we obtain the following equation:
\begin{equation}\label{10}
\left[\rho\frac{d^2}{d\rho^2}+\frac{1}{\rho}\frac{d}{d\rho}-\frac{(J_z-\frac{u}{2})^2}{4\rho}-\frac{\rho}{4}+E_u\right]\psi_u (\rho)=0,
\end{equation}
where 
\begin{equation}\label{EBARRA}
E_u=\frac{1}{4E\omega}\left[(E-g^0)^2-\left(p_z-\frac{\omega}{2}\right)^2+2\omega E\left(J_z+\frac{u}{2}\right)\right].
\end{equation}

Now, analyzing the asymptotic behavior/limit of Eq. \eqref{10} for $\rho\to 0$ and $\rho\to\infty$ \cite{Bakke,Bakke2,Bakke3,Bakke4,Cuzinatto,O2,O3,O4,O5,O6,Andrade,Cunha,Bragança}, we obtain a regular solution in the form:
\begin{equation}\label{F}
\psi_u(\rho)=C_u e^{-\rho/2}\rho^{\vert J_z-u/2\vert/2} F_u(\rho),
\end{equation}
where $C_u$ are normalization constants and $F_u(\rho)$ are unknown functions to be determined. In particular, to be a physically allowed solution, $\psi_u (\rho)$ must satisfy two boundary conditions, i.e. vanish at the origin and at infinity, such as $\psi_u(\rho\to 0)=\psi_u (\rho\to\infty)=0$. So, replacing the solution \eqref{F} in \eqref{10}, we obtain a second-order differential equation for $F_u(\rho)$:
\begin{equation}\label{11}
\left[\rho\frac{d^2}{d\rho^2}+\left(\left\vert J_z-\frac{u}{2}\right\vert+1-\rho\right)\frac{d}{d\rho}-\left(\frac{\left\vert J_z-\frac{u}{2}\right\vert+1}{2}-E_u\right)\right]F_u(\rho)=0,
\end{equation}
or yet (``Dvornikov-type form'')
\begin{equation}\label{12}
\left[\rho\frac{d^2}{d\rho^2}+\left(\left\vert J_z-\frac{u}{2}\right\vert+1-\rho\right)\frac{d}{d\rho}+\kappa\right]F_u(\rho)=0, \ \ \kappa=\left(-\frac{\left\vert J_z-\frac{u}{2}\right\vert+1}{2}+E_u\right).
\end{equation}

According to Refs. \cite{Bakke,Bakke2,Bakke3,Bakke4,Cuzinatto,O2,O3,O4,O5,O6,O8,O9,O10,Andrade,Cunha,Bragança,Carvalho,Cuzinatto2,Castro}, the equation above is an associated Laguerre equation (or a confluent hypergeometric equation), whose solution are the associated Laguerre polynomials (or confluent hypergeometric functions). Consequently, the parameter $\kappa$ must be equal to a positive integer, that is, $\kappa=N$, where $N=0,1,2,\ldots$ is the radial quantum number. In fact, the correct thing to say would be that the last term of \eqref{11} is a non-positive integer, but since the minus sign has already been incorporated into $\kappa$, it amounts to the same thing/result. Therefore, from the condition $\kappa=N$ (quantization condition), we obtain the following complete energy levels for the Dirac neutrino:
\begin{equation}\label{spectrum2}
\left[E-2N_u\omega-g^0\right]^2=(2N_u\omega)^2+4N_u\omega g^0+\left(p_z-\frac{\omega}{2}\right)^2, 
\end{equation}
where we define
\begin{equation}\label{N}
N_u\equiv\left(N+\frac{1}{2}+\frac{\vert J_z-\frac{u}{2}\vert}{2}-\frac{(J_z+\frac{u}{2})}{2}\right)=\left(N+\frac{1-u}{2}+\frac{\vert J_z-\frac{u}{2}\vert}{2}-\frac{(J_z-\frac{u}{2})}{2}\right), 
\end{equation}
being $N_u=N_{eff}=N_{total}$ a type of ``effective or total quantum number'' (since it depends on all others), and is similar to that of Refs. \cite{O2,O3,O4,O5,O6,Bueno,Salazar,Oliveira}. On the other hand, we see clearly/explicitly that for $J_z>0$ (positive angular momentum) with $u=+1$ (component $\psi_3$), we obtain exactly the particular energy levels of Dvornikov’s paper \cite{Dvornikov}. Furthermore, it is important to mention that Dvornikov has another paper that has this same problem (the spectrum does not depend on $J_z$), that is, the particular spectrum given by \eqref{spectrum1} has also been obtained/worked on in another paper of his, given by \cite{Dvornikov2015} (i.e. \cite{Dvornikov2015} is a shortened version of \cite{Dvornikov}).

%-------------------------------------------------------------------------

\section{Final remarks}

In this comment, we obtain the complete energy levels for Dvornikov's paper \cite{Dvornikov}, that is, the energy levels dependent on two quantum numbers, namely, the radial quantum number (given by $N=0,1,2,\ldots$) and the angular quantum number (given by $J_z=\pm\frac{1}{2},\pm \frac{3}{2}, \pm \frac{5}{2},\ldots$). In particular, what motivated us to do this was the fact that the quantized energy levels for particles (fermions or bosons) in polar, cylindrical, or spherical coordinates depend on two quantum numbers: a radial quantum number and an angular quantum number. In view of this, the following question/doubt arose: Why do the energy levels in Dvornikov’s paper only depend on one quantum number? In other words, Where did the angular quantum number given by $J_z$ go? So, using Studenikin’s paper as a starting point (as well as others in the literature \cite{Bakke,Bakke2,Bakke3,Bakke4,Cuzinatto,O1,O2,O3,O4,O5,O6,Cunha,Bragança}), we write one of the equations from Dvornikov's paper \cite{Dvornikov} in a matrix form. Later, we use the four-component Dirac spinor and obtain a set/system of four coupled first-order differential equations. From the first two equations with $m\to 0$, we obtain a (compact) second-order differential equation for the last two spinor components (i.e. $\psi_3$ and $\psi_4$). So, solving this equation, we obtain the neutrino energy levels, which explicitly depend on both $N$ and $J_z$. Finally, we note that for $J_z>0$ (positive angular momentum) with $u=+1$ (component $\psi_3$), we obtain exactly the particular energy levels of Dvornikov’s paper \cite{Dvornikov}.
 
%-------------------------------------------------------------------------
\section*{Acknowledgments}

\hspace{0.5cm}The author would like to thank the Conselho Nacional de Desenvolvimento Cient\'{\i}fico e Tecnol\'{o}gico (CNPq) for financial support through the postdoc grant No. 175392/2023-4, and also to the Department of Physics at the Universidade Federal da Para\'{i}ba (UFPB) for hospitality and support.

%-------------------------------------------------------------------------
\section*{Data availability statement}

\hspace{0.5cm} This manuscript has no associated data or the data will not be deposited. [Author’ comment: There is no data associated with this manuscript or no data has been used to prepare it.]


\begin{thebibliography}{99}
\section*{References}

\bibitem{Dvornikov} M. Dvornikov, J. High Energy Phys. {\bf 2014}, 1-15 (2014). {\color{blue}\url{https://doi.org/10.1007/JHEP10(2014)053}. {\href{https://arxiv.org/abs/1408.2735}{arXiv:1408.2735}}}

\bibitem{Bakke} K. Bakke, Gen. Relativ. Gravit. {\bf 45}, 1847–1859 (2013). {\color{blue}\url{https://doi.org/10.1007/s10714-013-1561-6}. {\href{https://arxiv.org/abs/1307.2847}{arXiv:1307.2847}}}

\bibitem{Bakke2} K. Bakke, and C. Furtado, Phys. Rev. D {\bf 80}, 024033 (2009). {\color{blue}\url{https://doi.org/10.1103/PhysRevD.82.084025}}.

\bibitem{Bakke3} K. Bakke, and C. Furtado, Phys. Rev. D {\bf 82}, 084025 (2010). {\color{blue}\url{https://doi.org/10.1103/PhysRevD.80.024033}}.

\bibitem{Bakke4} K. Bakke, and C. Furtado, Ann. Phys. (N.Y.) {\bf 336}, 489-504 (2013). {\color{blue}\url{https://doi.org/10.1016/j.aop.2013.06.007}. {\href{https://arxiv.org/abs/1307.2888}{arXiv:1307.2888}}}

\bibitem{Cuzinatto} R. R. Cuzinatto, M. de Montigny, and P. J. Pompeia, Gen. Relativ. Gravit. {\bf 51}, 107 (2019). {\color{blue}\url{https://doi.org/10.1007/s10714-019-2593-3}. {\href{https://arxiv.org/abs/1909.00904}{arXiv:1909.00904}}}

\bibitem{O1} R. R. S. Oliveira, Gen. Relativ. Gravit. {\bf 51}, 120 (2019). {\color{blue}\url{https://doi.org/10.1007/s10714-019-2606-2}. {\href{https://arxiv.org/abs/1907.00054}{arXiv:1907.00054}}}

\bibitem{O2} R. R. S. Oliveira, Eur. Phys. J. C {\bf 79}, 725 (2019). {\color{blue}\url{https://doi.org/10.1140/epjc/s10052-019-7237-y}}

\bibitem{O3} R. R. S. Oliveira, Gen. Relativ. Gravit. 52, {\bf 88} (2020). {\color{blue}\url{https://doi.org/10.1007/s10714-020-02743-6}. {\href{https://arxiv.org/abs/1906.07369}{arXiv:1906.07369}}}

\bibitem{O4} R. R. S. Oliveira, G. Alencar, and R. R. Landim, Gen. Relativ. Gravit. {\bf 55}, 15 (2023). {\color{blue}\url{https://doi.org/10.1007/s10714-022-03057-5}. {\href{https://arxiv.org/abs/2204.06057}{arXiv:2204.06057}}}

\bibitem{O5} R. R. S. Oliveira, Gen. Relativ. Gravit. {\bf 56}, 30 (2024). {\color{blue}\url{https://doi.org/10.1007/s10714-024-03209-9}. {\href{https://arxiv.org/abs/2402.15720}{arXiv:2402.15720}}}

\bibitem{O6} R. R. S. Oliveira, Class. Quantum Grav. {\bf 41}, 175017 (2024).
{\color{blue}\url{https://doi.org/10.1088/1361-6382/ad69f5}. {\href{https://arxiv.org/abs/2405.11334}{arXiv:2405.11334}}}

\bibitem{Andrade} F. M. Andrade, and E. O. Silva, Eur. Phys. J. C {\bf 74}, 3187 (2014). {\color{blue}\url{https://doi.org/10.1140/epjc/s10052-014-3187-6}. {\href{https://arxiv.org/abs/1403.4113}{arXiv:1403.4113}}}

\bibitem{Cunha} M. M. Cunha, H. S. Dias, and E. O. Silva, Phys. Rev. D {\bf 102}, 105020 (2020). {\color{blue}\url{https://doi.org/10.1103/PhysRevD.102.105020}. {\href{https://arxiv.org/abs/2007.08699}{arXiv:2007.08699}}}

\bibitem{Bragança} E. A. F. Bragança, R. L. L. Vitória, H. Belich, and E. B. de Mello, Eur. Phys. J. C {\bf 80}, 1-11 (2020). {\color{blue}\url{https://doi.org/10.1140/epjc/s10052-020-7774-4}}.

\bibitem{Carvalho} J. Carvalho, A. M. M. Carvalho, E. Cavalcante, and C. Furtado, Eur. Phys. J. C {\bf 76}, 1-9 (2016).  {\color{blue}\url{https://doi.org/10.1140/epjc/s10052-016-4189-3}. {\href{https://arxiv.org/abs/1603.06292}{arXiv:1603.06292}}}

\bibitem{Cuzinatto2} R. R. Cuzinatto, M. de Montigny, and P. J. Pompeia, Class. Quantum Grav. {\bf 39}, 075007 (2022). {\color{blue}\url{https://doi.org/10.1088/1361-6382/ac51bc}. {\href{https://arxiv.org/abs/2204.03049}{arXiv:2204.03049}}}

\bibitem{Castro} L. B. Castro, A. E. Obispo, and A. G. Jirón, Eur. Phys. J. C {\bf 84}, 536 (2024). {\color{blue}\url{https://doi.org/10.1140/epjc/s10052-024-12911-6}. {\href{https://arxiv.org/abs/2405.09471}{arXiv:2405.09471}}}

\bibitem{Studenikin} A. I. Studenikin, J. Phys. A: Math. Theor. {\bf 41}, 164047 (2008). {\color{blue}\url{https://doi.org/10.1088/1751-8113/41/16/164047}. {\href{https://arxiv.org/abs/0804.1417}{arXiv:0804.1417}}}

\bibitem{Dvornikov2} M. Dvornikov, and A. Studenikin, Phys. Rev. D {\bf 69}, 073001 (2004). {\color{blue}\url{https://doi.org/10.1103/PhysRevD.69.073001}. {\href{https://arxiv.org/abs/hep-ph/0305206}{arXiv:hep-ph/0305206}}}

\bibitem{Dvornikov3} M. Dvornikov, and A. Studenikin, High Energy Phys. {\bf 2002}, 016 (2002). {\color{blue}\url{https://doi.org/10.1088/1126-6708/2002/09/016}. {\href{https://arxiv.org/abs/hep-ph/0202113}{arXiv:hep-ph/0202113}}}

\bibitem{Dvornikov4} M. S. Dvornikov, and A. I. Studenikin, J. Exp. Theor. Phys. {\bf 99}, 254-269 (2004). {\color{blue}\url{https://doi.org/10.1134/1.1800181}. {\href{https://arxiv.org/abs/hep-ph/0411085}{arXiv:hep-ph/0411085}}}

\bibitem{Dvornikov5} M. Dvornikov, A. Grigoriev, and A. Studenikin, Int. J. Mod. Phys. D {\bf 14}, 309-321 (2005). {\color{blue}\url{https://doi.org/10.1142/S0218271805006018}. {\href{https://arxiv.org/abs/hep-ph/0406114}{arXiv:hep-ph/0406114}}}

\bibitem{Gavrilov1} S. P. Gavrilov, and D. M. Gitman, Phys. Rev. D {\bf 53}, 7162 (1996). {\color{blue}\url{https://doi.org/10.1103/PhysRevD.53.7162}. {\href{https://arxiv.org/abs/hep-th/9603152}{arXiv:hep-th/9603152}}}

\bibitem{Gavrilov2} S. P. Gavrilov, D. M. Gitman, and N. Yokomizo, Phys. Rev. D {\bf 86}, 125022 (2012). {\color{blue}\url{https://doi.org/10.1103/PhysRevD.86.125022}. {\href{https://arxiv.org/abs/1207.1749}{arXiv:1207.1749}}}

\bibitem{Vakarchuk} I. O. Vakarchuk, J. Phys. A: Math. Gen. {\bf 38}, 4727 (2005). {\color{blue}\url{https://doi.org/10.1088/0305-4470/38/21/016}. {\href{https://arxiv.org/abs/quant-ph/0502105}{arXiv:quant-ph/0502105}}}

\bibitem{Peres} N. M. R. Peres, A. H. Castro Neto, and F. Guinea, Phys. Rev. B {\bf 73}, 241403 (2006). {\color{blue}\url{https://doi.org/10.1103/PhysRevB.73.241403}. {\href{https://arxiv.org/abs/cond-mat/0603771}{arXiv:cond-mat/0603771}}}

\bibitem{O7} R. R. S. Oliveira, R. V. Maluf, and C. A. S. Almeida, Ann. Phys. (N.Y.) {\bf 400}, 1-8 (2019). {\color{blue}\url{https://doi.org/10.1016/j.aop.2018.11.005}. {\href{https://arxiv.org/abs/1809.03801}{arXiv:1809.03801}}}

\bibitem{O8} R. R. S. Oliveira, A. A. Araújo Filho, R. V. Maluf, and C. A. S. Almeida, J. Phys. A: Math. Theor. {\bf 53}, 045304 (2020). {\color{blue}\url{https://doi.org/10.1088/1751-8121/ab5cfb}. {\href{https://arxiv.org/abs/1812.07756}{arXiv:1812.07756}}}

\bibitem{O9} R. R. S. Oliveira, A. A. Araújo Filho, R. V. Maluf, and C. A. S. Almeida, J. Phys. A: Math. Theor. {\bf 54}, 028002 (2020). {\color{blue}\url{https://doi.org/10.1088/1751-8121/abd154}. {\href{https://arxiv.org/abs/2012.14282}{arXiv:2012.14282}}}

\bibitem{O10} R. R. S. Oliveira, G. Alencar, and R. R. Landim, Phys. Scr. {\bf 99}, 035226 (2024). {\color{blue}\url{https://doi.org/10.1088/1402-4896/ad25b3}. {\href{https://arxiv.org/abs/2211.09592}{arXiv:2211.09592}}}

\bibitem{Greiner} W. Greiner, {\it Relativistic quantum mechanics}, 3rd ed. (Springer, Berlin, 2000). 

\bibitem{Strange} P. Strange, {\it Relativistic Quantum Mechanics: with applications in condensed matter and atomic physics} (Cambridge: Cambridge University Press, 1998).

\bibitem{Bueno} M. J. Bueno, J. Lemos de Melo, C. Furtado, and A. M. M. Carvalho, Eur. Phys. J. Plus {\bf 129}, 201 (2014). {\color{blue}\url{https://doi.org/10.1140/epjp/i2014-14201-5}}.

\bibitem{Salazar} M. Salazar-Ramírez, D. Ojeda-Guillén, A. Morales-González, and V. H. García-Ortega, Eur. Phys. J. Plus {\bf 134}, 8 (2019). {\color{blue}\url{https://doi.org/10.1140/epjp/i2019-12381-0}}.

\bibitem{Oliveira} J. R. S. Oliveira, G. Q. Garcia, C. Furtado, and S. Sergeenkov, Ann. Phys. (N.Y.) {\bf 383}, 610-619 (2017). {\color{blue}\url{https://doi.org/10.1016/j.aop.2017.06.011}}.

\bibitem{Dvornikov2015} M. Dvornikov, Mod. Phys. Lett. A. {\bf 30}, 1530017 (2015). {\color{blue}\url{https://doi.org/10.1142/S0217732315300177}. {\href{https://arxiv.org/abs/1503.01431}{arXiv:1503.01431}}}

\end{thebibliography}
\end{document}